\documentclass[namedreferences]{SolarPhysics}
\usepackage[optionalrh]{spr-sola-addons}
\usepackage{graphicx}
\def\mc{meridional circulation}
\def\bl{Babcock--Leighton}
\def\Rs{R_{\odot}}

\begin{document}
\begin{opening}

\title{Quenching of Meridional Circulation in Flux Transport Dynamo Models}

\author{Bidya Binay \surname{Karak} Arnab Rai \surname{Choudhuri}}
\institute{Department~of~Physics,~Indian~Institute~of~Science,~Bangalore~560012,~India \email{arnab@physics.iisc.ernet.in}}
\runningtitle{Quenching of Meridional Circulation}
\runningauthor{Karak and Choudhuri}

\begin{abstract}
Guided by the recent observational result that the \mc\ of the Sun becomes weaker at
the time of the sunspot maximum, we have included a parametric quenching of the
meridional circulation in solar dynamo models such that the \mc\ becomes
weaker when the magnetic field at the base of the convection zone is
stronger. We find that a flux transport solar dynamo tends to become unstable
on including this quenching of \mc\ if the diffusivity in the convection
zone is less than about $2 \times 10^{11}$ cm$^2$ s$^{-1}$. The quenching 
of $\alpha$, however, has a stabilizing effect and it is possible to stabilize a dynamo
with low diffusivity with sufficiently strong $\alpha$-quenching.  For dynamo
models with high diffusivity, the quenching of \mc\ does not produce a large
effect and the dynamo remains stable. We present a solar-like solution from
a dynamo model with diffusivity $2.8 \times 10^{12}$ cm$^2$ s$^{-1}$ in which
the quenching of \mc\ makes the \mc\ vary periodically with solar cycle as observed
and does not have any other significant effect on the dynamo.
\end{abstract}
\end{opening}

\section{Introduction}
Flux transport dynamo model \cite{csd95, d95, dc99, nc02, cnc04, c10, ch11} is the most promising model 
of solar cycle at present. In this model, the strong toroidal 
field is generated near the base of the convection zone due to 
the stretching of the poloidal field by the strong differential rotation and 
the poloidal field is generated near the surface through the \bl\ mechanism. 
The \mc\ and the turbulent diffusivity are the two important flux transport 
agents from the source of the poloidal field (near the surface) to
the source region of the toroidal field (bottom of the
convection zone). The \mc\ plays a very crucial role in the flux transport
dynamo model by determining its period \cite{dc99} and also has an effect 
on the cycle strength \cite{ynm08, k10}. It appears that 
the \mc\ is crucial in modeling many irregular features of the solar cycle 
like the Waldmeier effect \cite{kc11} and the Maunder-like grand minima 
\cite{k10}. Moreover, Passos and Lopes (2008; see also Lopes and Passos 2009a) 
found the importance of \mc\ when they studied the solar cycle using 
a low-order dynamo model.

Our observational knowledge of the \mc\ is very limited at the present
time. It is poleward near the surface with an amplitude of around 
$20$ m~s$^{-1}$. However, the detailed profile of the \mc\ 
and the return flow at the bottom of the convection zone
are not observationally established yet. It is believed that the \mc\ is
caused by the combined effect 
of buoyancy, Reynolds stresses, latitudinal pressure gradient
and the Coriolis force acting on the mean zonal flow (Kitchatinov and R\"udiger 1995). 
Therefore we expect that there may be random variations in this flow due to 
fluctuations in any of these driving forces. From the periods of
various past cycles, Karak and Choudhuri (2011) have attempted
to draw some conclusions about the random variations in the
\mc\ (also see Passos and Lopes 2008; Lopes and Passos 2009b). 
In addition to these random variations, 
there must also be a periodic variation with the solar cycle due to the 
feedback of the Lorentz force of the dynamo-generated magnetic field. 
This type of periodic variation has recently been reported by 
Hathaway and Rightmire (2010) and Basu and Antia (2010), who found that 
the circulation speed was slightly weaker 
at the time of the sunspot maximum. Such a thing was also seen 
in a simulation of flux transport dynamo 
\cite{r06}. The main aim of this paper is to study the effects
of the periodic variation of \mc\ on the flux transport
dynamo model. We produce the periodic variation of \mc\ by
introducing a simple quenching of \mc\ by the dynamo-generated 
magnetic field. Our main conclusion is that the quenching of
the \mc\ has very different effects on flux transport dynamo models using 
different values of the turbulent diffusivity.  

During the last few years, two classes of flux transport dynamo models
have been developed in considerable details: (i) the high-diffusivity model  
in which the value of the turbulent diffusivity $\eta$ in the 
convection zone is taken in the range $\sim 10^{12}$--$10^{13}$ cm$^2$ s$^{-1}$
(Chatterjee, Nandy and Choudhuri 2004; Choudhuri, Chatterjee and Jiang 2007)
and (ii) the low-diffusivity model in which $\eta$ in the convection zone
is $\sim 10^{10}$--$10^{11}$ cm$^2$ s$^{-1}$ (Dikpati and Charbonneau 1999;
Dikpati and Gilman 2006).  We may note that the simple mixing length theory 
gives the value of $\eta$ 
$\sim 1$--$4 \times 10^{12}$ cm$^2$ s$^{-1}$ (Parker 1979, p. 629), similar
to what is used in the high-diffusivity model. We find that the quenching
of meridional circulation does not have any significant effect on the dynamo
when the diffusivity is high.  However, when the diffusivity is low, the dynamo model becomes
unstable on introducing the quenching of \mc\, unless we also introduce
a strong quenching of $\alpha$ to stabilize the solution. We show that
the physics of what is happening can be understood on the basis of
some results presented by Yeates, Nandy and 
Mackay (2008). It may be mentioned that several authors have recently
argued in favour of the high-diffusivity model (Chatterjee and Choudhuri
2006; Jiang, Chatterjee and Choudhuri 2007; Goel and Choudhuri 2009; 
Choudhuri and Karak 2009;  Hotta and Yokoyama 2010a, 2010b; Karak 2010;
Karak and Choudhuri 2011; Kitchatinov and Olemskoy 2011). The results of this paper provide further
support to the high-diffusivity model.

\section{Model}
In the flux transport dynamo model, the evolution of magnetic fields are 
governed by the following two equations:
\begin{equation}
\frac{\partial A}{\partial t} + \frac{1}{s}({\bf v}.\nabla)(s A)
= \eta_{p} \left( \nabla^2 - \frac{1}{s^2} \right) A + S(r, \theta; B),
\end{equation}

\begin{equation}
\frac{\partial B}{\partial t}
+ \frac{1}{r} \left[ \frac{\partial}{\partial r}
(r v_r B) + \frac{\partial}{\partial \theta}(v_{\theta} B) \right]
= \eta_{t} \left( \nabla^2 - \frac{1}{s^2} \right) B 
+ s({\bf B}_p.{\bf \nabla})\Omega + \frac{1}{r}\frac{d\eta_t}{dr}\frac{\partial{(rB)}}{\partial{r}},
\end{equation}
where $s = r \sin \theta$. 
Here  ${\bf B_p}$ correspond to the poloidal components of magnetic field which is given by 
the curl of $A(r, \theta, t) {\bf e}_{\phi}$, whereas $B (r, \theta, t)$ correspond to the
toroidal components of magnetic field. ${\bf v}$ is the 
velocity of the meridional flow. The term $S(r, \theta; B)$ is the responsible
for generating the poloidal field, whereas $\Omega$ is the internal angular velocity of
the Sun, and $\eta_p$, $\eta_t$ are the turbulent diffusivities for the poloidal
and toroidal fields respectively. 

For a particular solar dynamo model, we need to specify the values
of the various parameters, such as $\Omega$, ${\bf v}$, $\eta_p$, $\eta_t$
and $S(r, \theta; B)$.  Some of our calculations are done by using
exactly same parameters as used by Dikpati and Charbonneau (1999) in their `reference solution' 
except for the value of $u_0$ which is taken as $20$ m~s$^{-1}$ rather 
than $10$ m~s$^{-1}$ as quoted by them. Only with this value of $u_0$, we
are able to reproduce their results (Jiang, Chatterjee, and Choudhuri 2007).
To see the effect of quenching of \mc\ on this model, we introduce
the quenching as described in Section~3. We also have to suppress the
$\alpha$-quenching to some extent, in order to see the effect of
\mc\ quenching clearly, as discussed in Section~4. One of the advantages
of this model is that it gives periodic solutions even when the
diffusivity is increased by a few orders. To understand how the
results change on changing the diffusivity, we do runs with different
diffusivity while keeping most of the other things constant. Note that 
in this model a single value of magnetic turbulent diffusivity for 
the toroidal and the poloidal field is used. We call this
dynamo model the DC99 in spite of some differences with the original
`reference solution' of Dikpati and Charbonneau (1999). Our main
conclusion in \S4 will be that the quenching of \mc\ can make this
model unstable if the diffusivity in the solar convection zone is
assumed to be lower than about $2 \times 10^{11}$ cm$^2$ s$^{-1}$,
provided we do not have a strong $\alpha$-quenching to stabilize the
system.

One solar dynamo model which has been developed and used extensively
in our group is the model of 
Chatterjee, Nandy and Choudhuri (2004), who presented what they called
their `standard model'. We, however, recently used slightly 
different values of a few parameters which are discussed in Karak (2010). 
We call this K10 model.  Karak (2010) had used a rather high diffusivity of
$2.8 \times 10^{12}$ cm$^2$ s$^{-1}$. If we change the diffusivity to a
much lower value keeping the other parameters of this model constant,
we find that this model does not give periodic solutions. So we present
results of the \mc\ quenching on the K10 model only for the high diffusivity
and the model is found to be stable like the high-diffusivity version
of the DC99 model. 

\def\Ba\overline{B}

\section{Quenching of Meridional Circulation}

We have reliable data of the temporal variation of the surface \mc\ of the last 
one cycle only (Hathaway and Rightmire 2010; Basu and Antia 2010). 
Therefore, in most of our earlier kinematic calculations, we had used 
a constant value of \mc. It is easy to show that a toroidal
magnetic field at the base of the convection zone will have a poleward
Lorentz force (van Ballegooijen and Choudhuri 1988). 
During the solar maximum, when the toroidal field
at the base of the convection zone is particularly strong, we expect 
that the poleward Lorentz force due to it will oppose the \mc\ which is 
equatorward there and thereby would lead to a reduction in the \mc\
at the time of the solar maximum. One needs to consider this feedback of 
the magnetic field on the \mc\ while studying the solar cycle using 
kinematic dynamo models. For a full treatment of the problem, it is
necessary to solve the Navier--Stokes equation for the \mc\ and
to carefully address such questions as to how much time will
be required for the \mc\ to slow down at the surface after the poleward
Lorentz force starts acting at the base of the convection zone. We
are carrying out a detailed investigation of some of these issues,
which will be presented in a future paper.  In this paper, we present
results of dynamo calculations obtained with the simplistic
assumption that \mc\ throughout the convection zone gets reduced 
at the solar maximum because of a quenching due to the magnetic field. 
Therefore we introduce a quenching in the amplitude of the \mc\ 
in the following form:
\begin{equation}
v_0 =  v_0^\prime/[1+(\overline{B} /B'_0)^2]
\end{equation}
where $\overline{B}$ is the average toroidal field 
in the tachocline ($ r=0.65\Rs$ to $0.71\Rs$). 
Although the above form of the quenching seems very simple, it does a very 
good job in reducing the amplitude of \mc\ $v_0$ rapidly once the toroidal field 
approaches the value $B'_0$ and produces some striking effects on the behaviour
of the flux transport dynamo---especially if the diffusivity is low. We may also mention 
that although the \mc\ and the angular velocity are related, we are not considering
any quenching on the angular velocity in this work.

\section{Results}

We study the effect of \mc\ quenching on both the DC99 and K10 models
as mentioned in Section~2. For both the models, we first run the code for several solar 
cycles without the quenching in the \mc. Then we stop the code at some point
and introduce the quenching on meridional circulation 
using Equation (3). After this change in \mc, we run the code again for several 
solar cycles. For the DC99 model, we are able to do calculations by
varying diffusivity over a wide range.  For the K10 model, however, we
present results only for high diffusivity, since this model does not give
periodic solutions for low diffusivity.

\subsection{Results from the DC99 model}

As we pointed out
in Section~2, we use the model of Dikpati and Charbonneau (1999) apart from changing
the value of $u_0$ they quoted.  We make another crucial change. It is known
for a long time that the $\alpha$-quenching has a stabilizing effect on the
dynamo---see Section~1 of Choudhuri (1992). Dikpati and
Charbonneau (1999) had used a rather strong $\alpha$-quenching
given by
$$\alpha =  \alpha_0/[1+(B /B_0)^2]. \eqno(4)$$
This makes
the source function $S(r, \theta; B) = \alpha B$ in (1) vary as a function of $B$ as
shown by the solid line in Figure~1.  Such a strong $\alpha$-quenching would
suppress any possible instability induced by the quenching of \mc. In 
order to see the effects of \mc\ quenching, it is essential to make the
$\alpha$-quenching weaker. We present
calculations obtained with a milder $\alpha$-quenching 
given by
$$\alpha =  \alpha_0/[1+(|B| /B_0)^{1.1}]. \eqno(5)$$
and shown by the dashed
line in Figure~1. Without any $\alpha$-quenching at all, $S(r, \theta; B)$
which is equal to $\alpha B$ would simply increase linearly with $B$. 
We shall give arguments below why we
consider the $\alpha$-quenching used by Dikpati and Charbonneau (1999)
to be unrealistically strong. It may be noted that, before we introduce
the quenching in \mc, the $\alpha$-quenching is the only source of nonlinearity
in the problem and determines the amplitude of the magnetic field. If
we increase the numerical value of $B_0$ in the code by a factor $s$, the
magnetic field in the solution everywhere increases uniformly by the same
factor $s$.

\begin{figure}
\centering
\centering{\includegraphics[width=9cm]{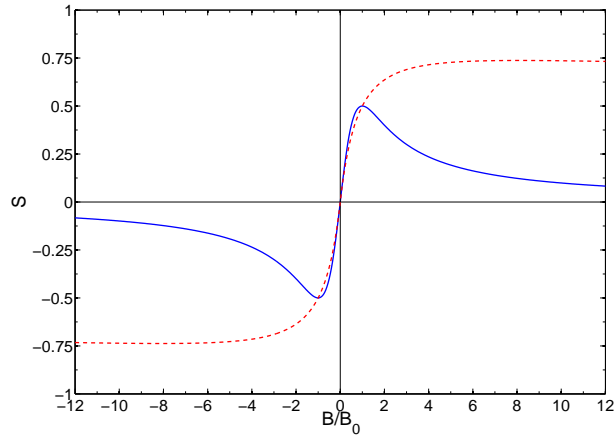}}
\caption{The poloidal field generation factor $S(r, \theta; B)$
as a function of the toroidal field. 
The quenching profile shown by the solid line was used in Dikpati and Charbonneau (1999). 
The profile shown by the dashed line is used in most of the calculations in this paper.
\label{alquench}}
\end{figure}
 
\begin{figure}
\centering{\includegraphics[width=14cm]{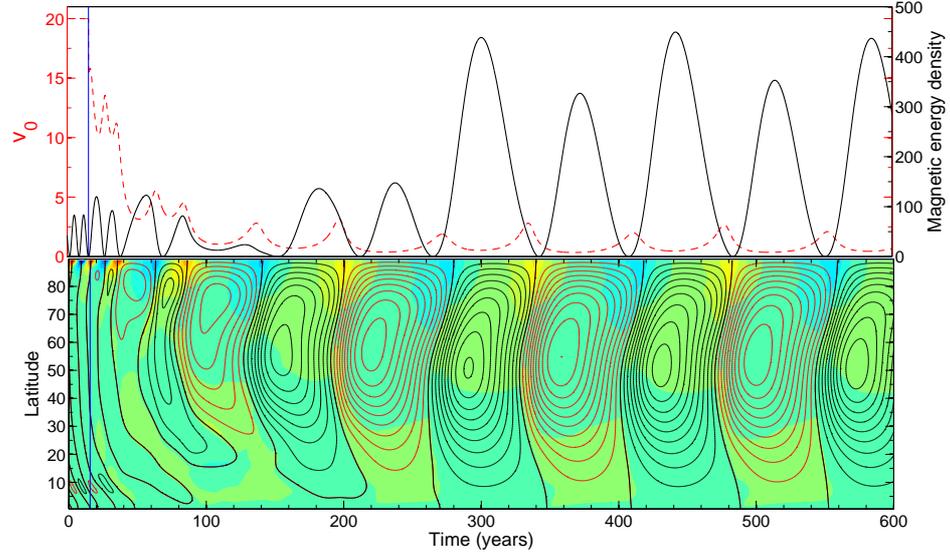}}
\vspace{-2.0cm}
\caption{Results from DC99 model: Upper panel shows the variation of the
amplitude of meridional circulation $v_0$ (dashed line) and the
magnetic energy density of the toroidal field at latitude $15^{\circ}$ at the
bottom of the convection zone (solid line).
The vertical solid line indicates the time of the
initiation of quenching.
Lower panel shows the butterfly diagram of the toroidal field (contours). 
The background shows the weak diffuse radial field on the solar surface.
The diffusivity used in this case is $5 \times 10^{10}$ cm$^2$ s$^{-1}$.
\label{bflow}}
\end{figure}

Figure~2 shows the results of \mc\ quenching on the DC99 model using the
same low diffusivity of $5 \times 10^{10}$ cm$^2$ s$^{-1}$
as used by Dikpati and Charbonneau (1999). 
We have used a rather weak quenching of \mc\ in which $B'_0$ appearing in (3)
is taken to be $B'_0/B_0 = 9.4$. Note that a larger value of $B'_0/B_0$
implies a weaker quenching of \mc.
The upper panel shows the variation of the $v_0$ (dashed line) and the 
magnetic energy density of the toroidal field at 
latitude $15^{\circ}$ at the bottom of the convection zone (solid line). 
The vertical solid line indicates the time when the quenching in \mc\ in 
accordance with (3) is included. We see that immediately 
after the commencement of the quenching the value of $v_0$ drops to 
around $16$ m~s$^{-1}$ from the usual value $20$ m~s$^{-1}$ and then 
it oscillates with the solar cycle. However, the value of $v_0$ keeps
decreasing with time and falls close to zero in about 
four to five solar cycles. Then it oscillates 
with the solar cycle remaining close to the zero value. We also
note that the cycle period increases as $v_0$ decreases, which is
expected in a flux transport dynamo. 
In the butterfly diagram given in the bottom panel, we see that the 
equatorial propagation of the toroidal field does not happen beyond a 
few solar cycles after the commencement of the quenching. This is because 
the \mc\ which is responsible for this equatorial propagation of the 
toroidal field \cite{csd95} has become very weak. It is clear that
the low-diffusivity model fails to give solar-like 
oscillation after the inclusion of the quenching in \mc.

Next, Figure~3 shows the results on increasing the diffusivity of the DC99 model
to the value $2.8 \times 10^{12}$ cm$^2$ s$^{-1}$ (which is used in the K10
model). When the diffusivity is increased, we need to increase the value
of $\alpha$ along with it, otherwise the solutions decay away.  While the
value of $\alpha_0$ used to generate Figure~2 was 
0.2 m~s$^{-1}$ the same as what was used in the
original work of DC99, we now take $\alpha_0$ to be 215 m~s$^{-1}$. Except the
changes in diffusivity and $\alpha_0$, all the other parameters remain the same
between Figures~2 and 3.  Figure~3 plots the same things
as Figure~2, for an increased diffusivity.  We find that now we have a stable
periodic solution, although the butterfly diagram does not look solar-like and
the period is very short. If we run the code without the quenching of meridional
circulation, then also the butterfly diagram looks very much like what we see
in Figure~3, indicating that this quenching does not have too much effect
when the diffusivity is high.  

\begin{figure}
\centering{\includegraphics[width=14cm]{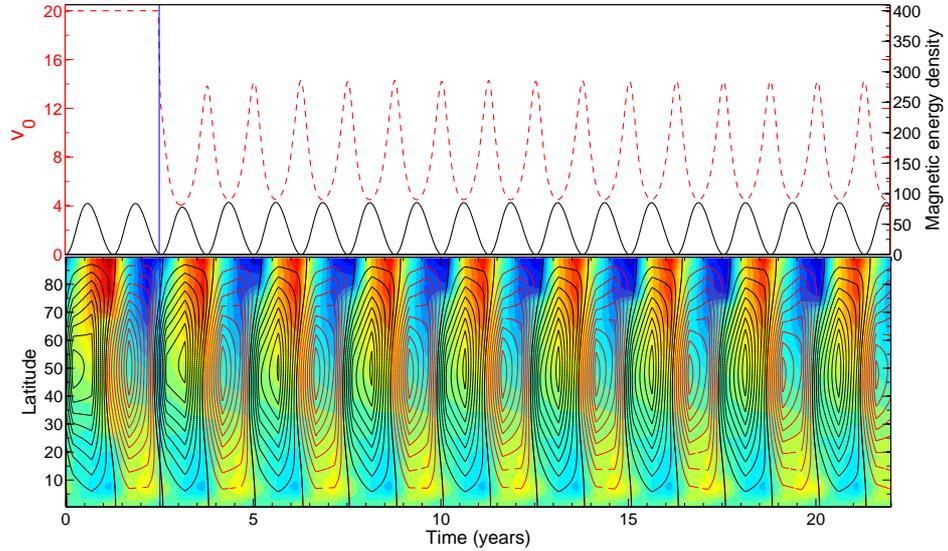}}
\vspace{-2.0cm}
\caption{Same as Figure~2 but in this case the diffusivity is increased
to $2.8 \times 10^{12}$ cm$^2$ s$^{-1}$.
\label{bflow}}
\end{figure}
\begin{figure}
\centering{\includegraphics[width=12cm]{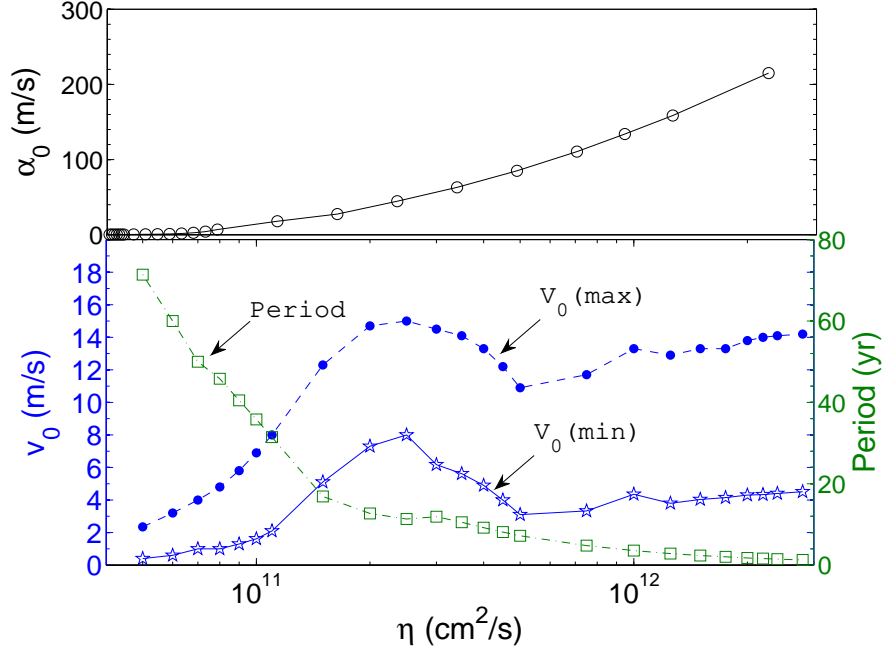}}
\vspace{-2.0cm}
\caption{Upper panel shows the values of $\alpha_0$ required to
get the stable solution when we change the diffusivity $\eta$ (shown along the horizontal axis).
Lower panel shows the variation of cycle period (square symbol), maximum $v_0$ (filled circle) 
and the minimum $v_0$ (star symbol) of the relaxed solutions as a function of diffusivity.
\label{parameter}}
\end{figure}

To understand how the nature of the solution changes from Figure~2 to Figure~3,
we have made runs for different values of diffusivity.  When increasing diffusivity,
we have also increased $\alpha_0$ in such a way that the toroidal field comes out to
have very similar strengths in all the runs. The upper panel of Figure~4 shows
the values of $\alpha_0$ chosen for different values of diffusivity.  The lower
panel of Figure~4 shows how other important properties of the solution change
of changing diffusivity.  One of the things shown is how the period of the
eventual relaxed solution changes on
changing the diffusivity.  For any diffusivity, the amplitude of meridional
circulation oscillates between a maximum and a minimum value, as we see in
Figures~2 and 3.  In the lower panel of Figure~4, we also plot the maximum and 
minimum values of the meridional circulation for different diffusivities.  
When the diffusivity is low, as in the case of Figure~2, the quenching of
meridional circulation makes the system unstable so that it eventually relaxes
to a solution in which the meridional circulation falls to a very low value
and the period is large.  On the other hand, when the diffusivity is high,
the solution has an asymptotic low value of period whereas meridional circulation
remains much stronger.  The maximum value of meridional circulation
in the high-diffusivity limit is essentially
the value that we would see when the quenching of meridional circulation
is switched off and the effect of quenching is not significant when 
diffusivity is high.  Figure~4 makes it clear that the transition from
instability to stability is not a very sharp transition, but takes place
roughly around the value $2\times 10^{11}$ cm$^2$~s$^{-1}$ of diffusivity.

The reason for this kind of peculiar behaviour can be understood from the
analysis of Yeates, Nandy, and Mackay (2008).  
When $v_0$ decreases due to the quenching, the cycle period is longer and
the poloidal field spends more time in the convection zone. This 
will result in two opposing effects. On the one hand, diffusion gets more
time to act on the poloidal field and the toroidal field ultimately
produced from this weaker poloidal field will tend to be weaker.
On the other hand, the differential rotation also gets more time to generate the toroidal 
field, thereby tending to make the toroidal field stronger. 
Whether the toroidal field will finally be weaker or stronger will depend 
on which of these two competing effects wins over. When the diffusivity 
is high, the diffusion of the fields is more important and the toroidal
field is weaker when the \mc\ decreases due to the quenching 
because diffusion has more time to act on the fields.  
This decrease of the toroidal field will make the \mc\ stronger according
to Equation (3). Thus the \mc\ will try to bounce back to a 
higher value. In this way, this model stabilizes under the quenching
of the \mc. However, if the diffusivity is low, things are opposite. 
When the \mc\ decreases due to  
quenching of the strong toroidal field and the dynamo period increases,  
the differential rotation generating the toroidal field for a longer
time is the dominant effect rather than diffusion.  Then the toroidal
field becomes stronger and this decreases the \mc\ further according
to Equation (3). This leads to a runaway unstable situation until the
\mc\ drops to very low values.  These arguments explain the results
seen in Figures~2 and 3. 

We also made some runs by making $B'_0/B_0$ lower than 9.4 used to
generate Figures 2--4, i.e.\ by making the quenching of \mc\ stronger.
Then the fall in the value of $v_0$ after introducing the quenching
was faster than what is seen in Figure~2 and the eventual periods were
longer.  Otherwise, the qualitative behaviour of the system does not
change when $B'_0/B_0$ is varied between 1 and 10.

As we pointed out, we have used a much weaker $\alpha$-quenching in
our calculations (as indicated by the dashed curve in Figure~1) than
what Dikpati and Charbonneau (1999) had used.  When we repeated our
calculations by using the stronger $\alpha$-quenching of Dikpati and
Charbonneau (1999) shown by the solid curve in Figure~1, we found that the
dynamo solutions do not change much on introducing the quenching
of the \mc\ even when the diffusivity is low.  
In other words, the strong $\alpha$-quenching suppresses 
the instability induced by the quenching of the \mc.  
It is not difficult to understand
physically what is happening.  The strong $\alpha$-quenching does not
allow the magnetic fields to grow the way they would grow in its
absence and thus the tendency of runaway growth of the toroidal field
on introducing the quenching of the \mc\ in the low-diffusivity case
is stabilized.

Since Figure 2 does not agree with the behaviours of solar magnetic
fields, one is tempted to conclude that the solar dynamo could
not be a low-diffusivity dynamo---unless there is a stabilizing
effect due to a strong $\alpha$-quenching.  So it is a crucial question
whether a strong $\alpha$-quenching is expected in the Sun. The
Babcock--Leighton mechanism, which is parameterized by $\alpha$,
depends on the tilts of active regions.  These tilts are produced
by the Coriolis force acting on the rising flux tubes and the effect
of the Coriolis force certainly does become weaker when the magnetic
field is stronger (Choudhuri 1989; D'Silva and Choudhuri 1993). 
However, magnetic fields of different sunspots do not vary too
much, suggesting that the toroidal field probably becomes buoyant
when it reaches a critical value and rising flux tubes within the
convection zone may not have widely different values of the magnetic
field.  More generation of toroidal field probably means more
active regions and not stronger magnetic fields within individual
active regions. If that is the case, then we do not think that one can
invoke a strong $\alpha$-quenching to stabilize the dynamo. In
spite of many uncertainties in our present understanding of the
solar dynamo and $\alpha$-quenching, it seems unlikely that
the solar dynamo is a low-diffusivity dynamo
which has a tendency of becoming unstable on introducing the quenching of the \mc.

\subsection{Results from K10 model} 

Let us now present the results from the K10 model, which uses
a high diffusivity $2.8 \times 10^{12}$ cm$^2$ s$^{-1}$ and produces
solar-like solutions.  We saw above that the DC model with such
a high diffusivity remains impervious to the effects of \mc\
quenching, but the solutions do not look solar-like.  Now we
find that the K10 model, which uses this high diffusivity and
produces solar-like solutions, also does not 
change qualitatively on introducing the quenching
of \mc.  As in the case of Figures~2 and 3, the dashed (red) line 
in Figure~5 shows the 
variation of $v_0$ whereas the solid (black) line shows the variation 
of the magnetic energy density of the toroidal field at 
latitude $15^{\circ}$ at the bottom of the convection zone 
(a measure of sunspot number). 
The vertical solid line indicates the time of introduction of the quenching.
We see from this figure that, soon after introducing 
the quenching, the value of $v_0$ drops to around $24$ m~s$^{-1}$ from the 
usual value $30$ m~s$^{-1}$. Then $v_0$ oscillates with the solar cycle, becoming 
weaker when the toroidal field becomes strong and vice versa. Since
the results obtained from the K10 model can be compared with observations, it 
may be noted that the amplitude variation of the $v_0$ is comparable to 
observational plot shown in Figure~4 of Hathaway and Rightmire (2010). 
We achieved this by suitably adjusting the parameter $B'_0$ appearing
in (3).  When the quenching is not present, the only nonlinearity
in our model comes from magnetic buoyancy, which is treated by
allowing the toroidal field to erupt whenever its value crosses
a critical value $B_c$ within the convection zone (Chatterjee,
Nandy and Choudhuri 2004).  This limits the growth of the dynamo
and makes the strongest toroidal fields at the bottom of the
convection zone hover around $B_c$.  It is $B_c$ which determines
the scale of the magnetic field in K10 model just as $B_0$ determines
the scale of the magnetic field in the DC99 model. We have used $B'_0/ B_c 
= 1.0/0.8$ which make the variation of the \mc\ comparable to what is observed.

\begin{figure}
\centering
\centering{\includegraphics[width=14cm]{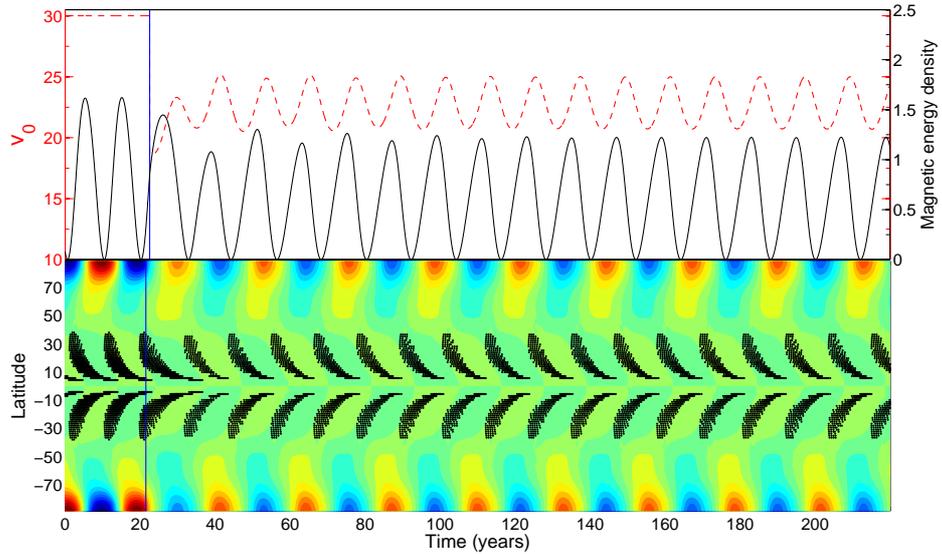}}
\vspace{-2.0cm}
\caption{Results from K10 model: Upper panel shows the variation of the amplitude of \mc\ $v_0$ (dashed line) and the
magnetic energy density of the toroidal field at latitude $15^{\circ}$ at the bottom of the convection zone
(solid line). The vertical solid line indicates the time of the initiation of quenching.
Lower panel shows the butterfly diagram of the sunspot eruptions.
The background shows diffuse radial field on the solar surface.
\label{bfhigh}}
\end{figure}

The bottom panel of Figure~5 shows the butterfly diagram of the sunspot eruptions 
along with the radial field on the surface.  We point out
that we have chosen the value of $v_0^\prime$ (which is same as $v_0$ in Karak 2010)
in such a way that after the addition of quenching 
the period of the solar cycle becomes close to 11~year.
We see that 
the butterfly diagram with the inclusion of the quenching of \mc\
looks very similar to the butterfly diagrams without such
quenching as presented by Chatterjee, Nandy and Choudhuri (2004).
We may point out that the high-diffusivity solution from the DC99
model shown in Figure~3, apart from not producing solar-like butterfly
diagrams, had a rather small period.  We believe that the main reason
behind this small period is that magnetic buoyancy was treated 
differently in this model.  See Choudhuri, Nandy, and Chatterjee (2005)
for a discussion of how the period of the dynamo can be very
different on treating magnetic buoyancy differently, even when
the other parameters are kept the same.

Finally, since we introduced stochastic fluctuations in the poloidal field generation 
process in some of our calculations, one last question we address is whether
dynamo models with stochastic fluctuations behave differently on introducing
the quenching of \mc. The result is presented in Figure~6. In this case, the amplitude of 
$\alpha$ is changed after the coherence time 6 months, the level of fluctuations
being $133\%$. We find that now there are some irregularities in the cycles as
we would expect and as we find in the calculations without including the
quenching in the \mc.

Our conclusion is that, in the case of the high-diffusivity K10 model, the results
remain qualitatively similar whether we include the quenching in the \mc\ or not.

\begin{figure}
\centering
\centering{\includegraphics[width=14cm]{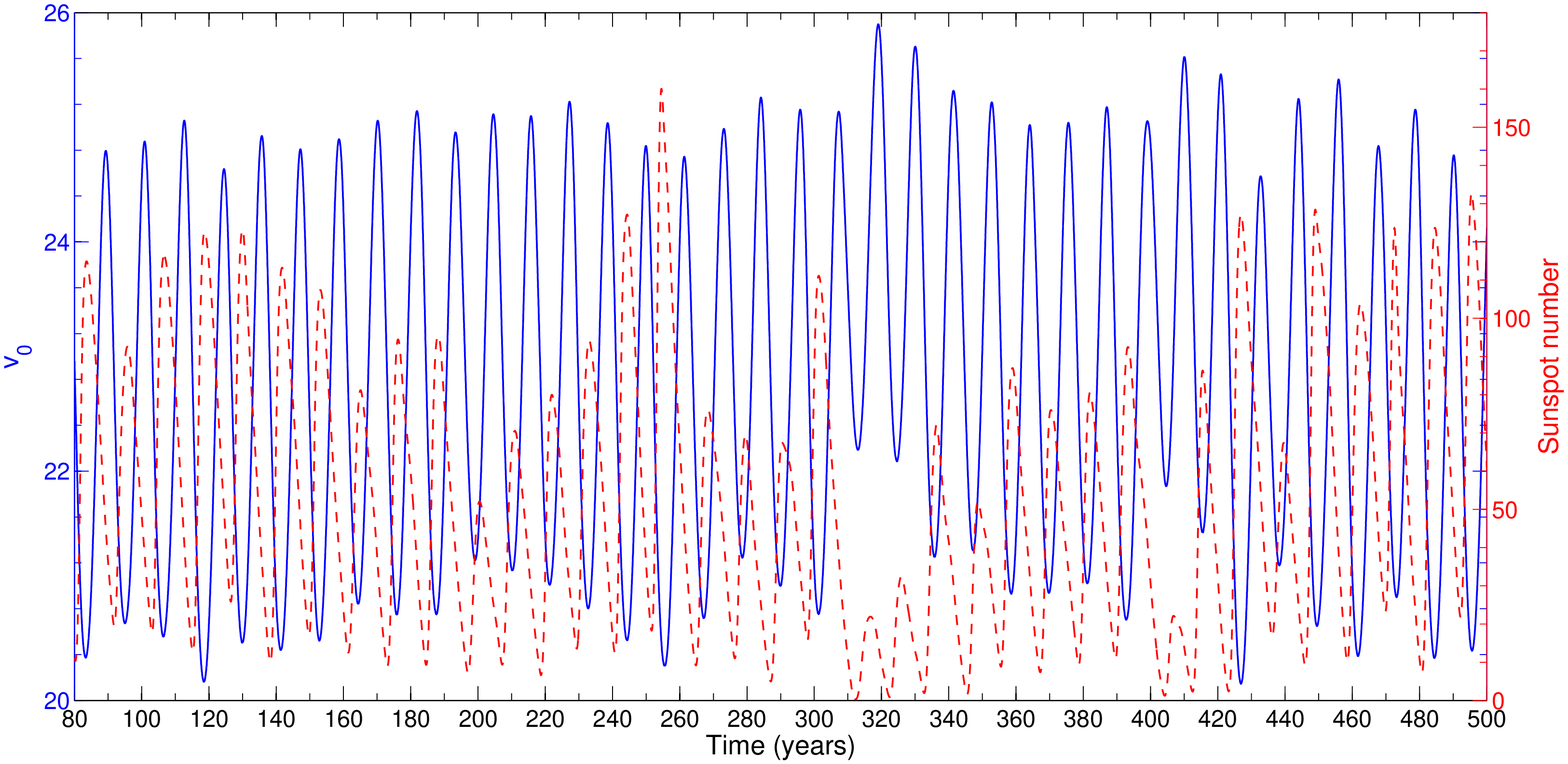}}
\caption{Results obtained by including stochastic fluctuations in the poloidal 
field generation process in the model of which results have been presented
in Figure~5. Note that in this case we have plotted the sunspot eruptions 
(dashed line) instead of energy density of toroidal field as we have done earlier.
The solid line shows the amplitude of \mc. 
\label{fluchigh}}
\end{figure}
 
\section{Conclusion}

We have considered the Lorentz feedback of the dynamo-generated magnetic 
field on the meridional circulation. We first performed numerical experiments 
on the DC99 dynamo model, varying the diffusivity from the rather
low value originally used by Dikpati and Charbonneau (1999) to values 
nearly two orders of magnitude larger. 
In the low-diffusivity situation, we found that the quenching of \mc\ 
leads to an instability which ultimately reduces the \mc\ to a very low value 
and the solution has a very large period. It is
true that this instability can be suppressed by a sufficiently strong
$\alpha$-quenching.  However, we argue that such a strong $\alpha$-quenching
is unphysical, making it unlikely that
the solar dynamo is a low-diffusivity 
dynamo. On the other hand, when we run the DC99 model with a high diffusivity,
we find it to be stable though the results do not look solar-like.  To show
the effect of the quenching of \mc\ on a solar-like dynamo model with high
diffusivity, we carry on some runs with the K10 model.  In this case also,
the solution is stable and the quenching of \mc\ does not have a big
effect on the dynamo except producing a period variation of the \mc\ as
observed.  Our main conclusion is that a solar dynamo model with
the quenching of the \mc\ becomes unstable if the diffusivity is
lower than about $2 \times 10^{11}$ cm$^2$ s$^{-1}$, unless
there is also a strong $\alpha$-quenching to suppress this
instability.  Since $\alpha$-quenching in the real Sun is unlikely
to be as strong as assumed by Dikpati and Charbonneau (1999), our
study indicates that the solar dynamo is most probably a high-diffusivity
dynamo. Several earlier authors already pointed out that the 
high-diffusivity model of the solar dynamo was much more successful
in explaining such things
as the parity of the solar magnetic fields (Chatterjee, Nandy
and Choudhuri 2004; Hotta and Yokoyama 2010b; Kitchatinov and Olemskoy 2011), 
the hemispheric coupling 
\cite{cc06, gc09}, the correlation between the polar field at the 
solar minimum and the sunspot number in the next cycle \cite{jcc07},
the Waldmeier effect (Karak and Choudhuri 2011), the
Maunder minimum \cite{ck09, k10}, the periods and the 
amplitudes of the last $23$ cycles \cite{k10}.  Our present study
further strengthens the case for the high-diffusivity dynamo.

The main effect of the quenching of \mc\ in a high-diffusivity dynamo
is that the \mc\ varies with the solar cycle in a periodic way, becoming
weaker at the time of the solar maximum.  We have chosen the parameters
of our model such that this periodic variation of the \mc\ matches with
what is reported by Hathaway and Rightmire (2010) from observations.
In a recent work, Nandy, Munoz-Jaramillo and Martens (2011) have assumed
that the \mc\ changes randomly at every solar maximum, without having 
any correlation with the strength of the cycle. We disagree
with this assumption and believe that the \mc\ decreases at the solar
maximum due to the Lorentz force of the magnetic fields in a systematic
deterministic way.  In this paper, we have treated the back-reaction
of the magnetic field on the \mc\ in a very simplistic way through Equation (3).
We have embarked on a more detailed calculation of this by solving
the Navier--Stokes equation and hope to present the results of this
calculation in a future paper.

\section*{Acknowledgment}
We would like to thank the anonymous referee for valuable suggestions which 
helped us to improve the manuscript.
This work is partly supported by DST through the J. C. Bose Fellowship awarded to ARC. 
BBK thanks CSIR, India for financial support.

\end{document}